# Band-Gap Engineering of Phononic Crystals:
# A Computational Survey of Two-Dimensional Systems

PSI-9039 SR-1307

E. A. Rietman and J. M. Glynn
Physical Sciences Inc.
Andover, Massachusetts 01810
ear@psicorp.com, jmg@psicorp.com

## Abstract

We present graphic results with high-levels of abstraction to describe the basic principles and rules of thumb for acoustic or phononic band-gap engineering. We use these rules for developing an improved machine mount for damping acoustic vibrations, a phononic lens, and a frequency selective filter in the acoustic regime.

## Introduction and Motivation

Following the publication of Joannopoulos et al (1995)[1] the field of photonic crystals has steadily grown. Parallel to that has been the field of phononic crystal research. Most of the work in both of these areas has been in proof of concept to improve theoretical understanding, or computational demonstration of simple devices. The band-gaps in both types of crystals are controlled by material composition, lattice spacing, crystal-packing arrangement, crystal orientation, and size of the elements comprising the crystal.

There have already been far too many papers on phononic crystals to review here, however, of particular interest for acoustic band-gap engineering are those by Garcia et al. (2003),[2] Kushwaha and Halevi (1994),[3] Lai et al. (2001),[4] Sigmund and Jensen (2003),[5] Caballero et al. (1999),[6] and Sliwa and Krawczyk (2005).[7]

Several novel devices have been described including negative refraction,[8-10] lenses,[11-13] waveguides,[14,15] and mode coupling and demultiplexing.[16,17]

The basis of phononic crystals dates back to Newton who imagined that sound waves propagated through air in the same way that an elastic wave would propagate along a lattice of point masses connected by springs with an elastic force constant E. This force constant is identical to the modulus of the material. Of course with phononic crystals of materials with differing modulus the calculations are a little more complicated than this simple model.

Based on Newton's observation that the elastic force constant is of key importance, we may deduce that a key factor for acoustic band-gap engineering is impedance mismatch between periodic elements comprising the crystal and the surrounding medium. When an advancing wave-front meets a material with very high impedance it will tend to increase its phase velocity through that medium. Likewise, when the advancing wave-front meets a low impedance medium it will slow down. We can exploit this concept with periodic (and handcrafted) arrangements of impedance mismatched elements to affect acoustic waves in the crystal – essentially band-gap engineering.

For inhomogenous solids the wave equation is given by

$$\frac{\partial^2 u_j^i}{\partial t^2} = \frac{1}{\rho_j}\left\{\frac{\partial}{\partial x_i}\left(\lambda \frac{\partial u_j^{\,i}}{\partial x_l}\right) + \frac{\partial}{\partial x_l}\left[\mu\left(\frac{\partial u_j^i}{\partial x_l} + \frac{\partial u_j^l}{\partial x_i}\right)\right]\right\}$$

where $u^i$ is the $i^{th}$ component displacement vector. The subscript j is in reference to the medium (medium 1 or medium 2); $\lambda$, $\mu$ are the Lame coefficients, $\rho$ is the density, and the longitudinal and transverse speed of sound are given by

$$c_l = \sqrt{(\lambda + 2\mu)/\rho}$$
$$c_t = \sqrt{\mu/\rho} \; .$$

The Lame coefficients can be expressed as Young's modulus *E*.

$$E_t = \rho c_t^2 = \mu$$
$$E_l = \rho c_l^2 = \lambda + 2\mu$$

Given the importance of Young's modulus to elastic vibrations in lattices we express our survey results in these terms. In this paper we describe a numerical survey of materials, lattice spacing, packing arrangements, and crystal orientations. From compiled graphical results we observe that as the Young's modulus increases, the width of the first (lowest frequency) band-gap also increases. This trend is observed for both cubic (X and M direction) and hexagonal crystals (K and M directions) at several filling fractions and rod diameters. After presenting our survey results, we describe several acoustic device ideas we have developed based on this understanding. We solved the wave equation with finite element modeling using Comsol Multiphysics® software.

## Computational Crystal Survey

We constructed (computationally) two-dimensional cubic and hexagonal crystals built from circular rods. We used three rod diameters: 3.175 mm, 6.35 mm and 9.525 mm (0.125, 0.25 and 0.375 inches respectively); and three filling fractions 0.90699, 0.403066, and 0.29613. All nine possible combinations of these rod diameters and filling fractions were investigated. For the cubic crystals we investigated X and M directions, and for hexagonal we investigated K and M directions. Thus, for each material combination we computed the acoustic properties for eighteen different crystals/orientations.

All the rods were embedded in urethane impedance match with water ($\rho = 1000$ kg/m$^3$; $c$ = 1497 m/sec). The materials we selected were alumina ($\rho = 3860$ kg/m$^3$; $c$ = 10520 m/sec; E=3.61X10$^{11}$ Pa), stainless steel ($\rho = 7850$ kg/m$^3$; $c$ = 5790 m/sec; E=1.03X10$^{11}$ Pa), aluminum ( $\rho = 2700\,\text{kg/m}^3$; $c$ = 6420 m/sec; E=6.9X$10^{10}$ Pa) and nylon ( $\rho = 1130\,\text{kg/m}^3$; $c$ = 2675 m/sec; E=2.4X$10^9$ Pa).

In addition we investigated X and M in cubic and K and M in hexagonal polyester ( $\rho = 1350\,\text{kg/m}^3$; $c$ = 2100 m/sec; E=4.41X$10^9$ Pa)) and graphite ( $\rho = 2200\,\text{kg/m}^3$; $c$ = 3310 m/sec; E=2.41X$10^{10}$ Pa) packed in urethane. These were done for all three rod diameters but only for filling fraction 0.29613.

Basic crystal systems were designed following the guide shown in Figure 1. In this figure the crystals is comprised of 3.175 mm rods in a hexagonal configuration and filling fraction 0.403. The basic crystal is about 3.5 cm X 5 cm surrounded by an urethane. To the left side of the crystal is an acoustic pressure source producing plane waves. All our studies were from 10 kHz to 200 kHz. About 2 cm on the right of the crystal is an imaginary box used for integration. In this region we integrated the acoustic energy for preparing the transmission spectra. The boundaries, except for the pressure source, are water impedance.

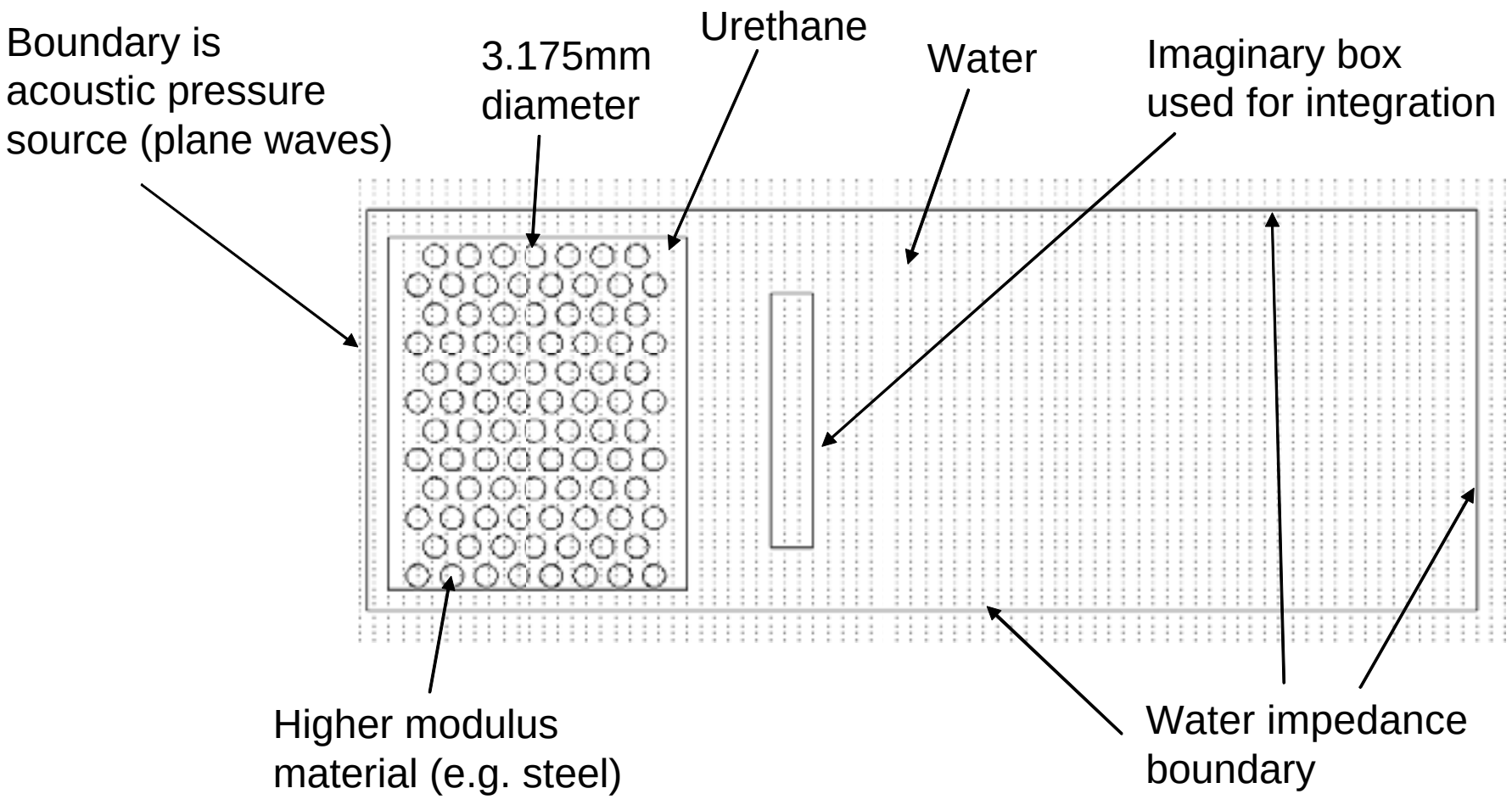

**Figure 1. Schematic diagram of one of the crystals used in this study. Note, water and urethane have the same acoustic impedance.**

The multi-dimensional survey of acoustic transmission for alumina, stainless steel, aluminum and nylon in urethane are shown in Figure 2 and displayed as a function of Young's modulus. The top half of the figure shows the spectra for cubic packing and the bottom half shows the spectra for hexagonal packing. Each material type is displayed as a set of 3 X 3 graphs. The 3 X 3 blocks of graphs are arrayed as filling fraction decreasing from top to bottom and rod diameter increasing from left to right. The abscissa on any of the small spectra is frequency from 10 kHz to 200 kHz. The ordinate is the transmitted energy in arbitrary units. Each of the small graphs contains the two crystal orientations we investigated (X, M cubic; K, M hexagonal). In each case the M direction is green.

Presenting the results this way clearly shows the six-dimensional data in a format where differences and similarities can be quickly appreciated. The first thing to notice is that the width and center frequency for the first band gap is clearly a function of the Young's modulus.
A second thing to notice is that the band gap is clearly related to the lattice spacing, a function of the filling fraction and the rod diameter. Lastly, we see that the band gaps for nylon are not as pronounced as the others because the modulus of nylon (E=2.4X$10^9$ Pa ) is nearing that of urethane (E=2.24X$10^9$ Pa).

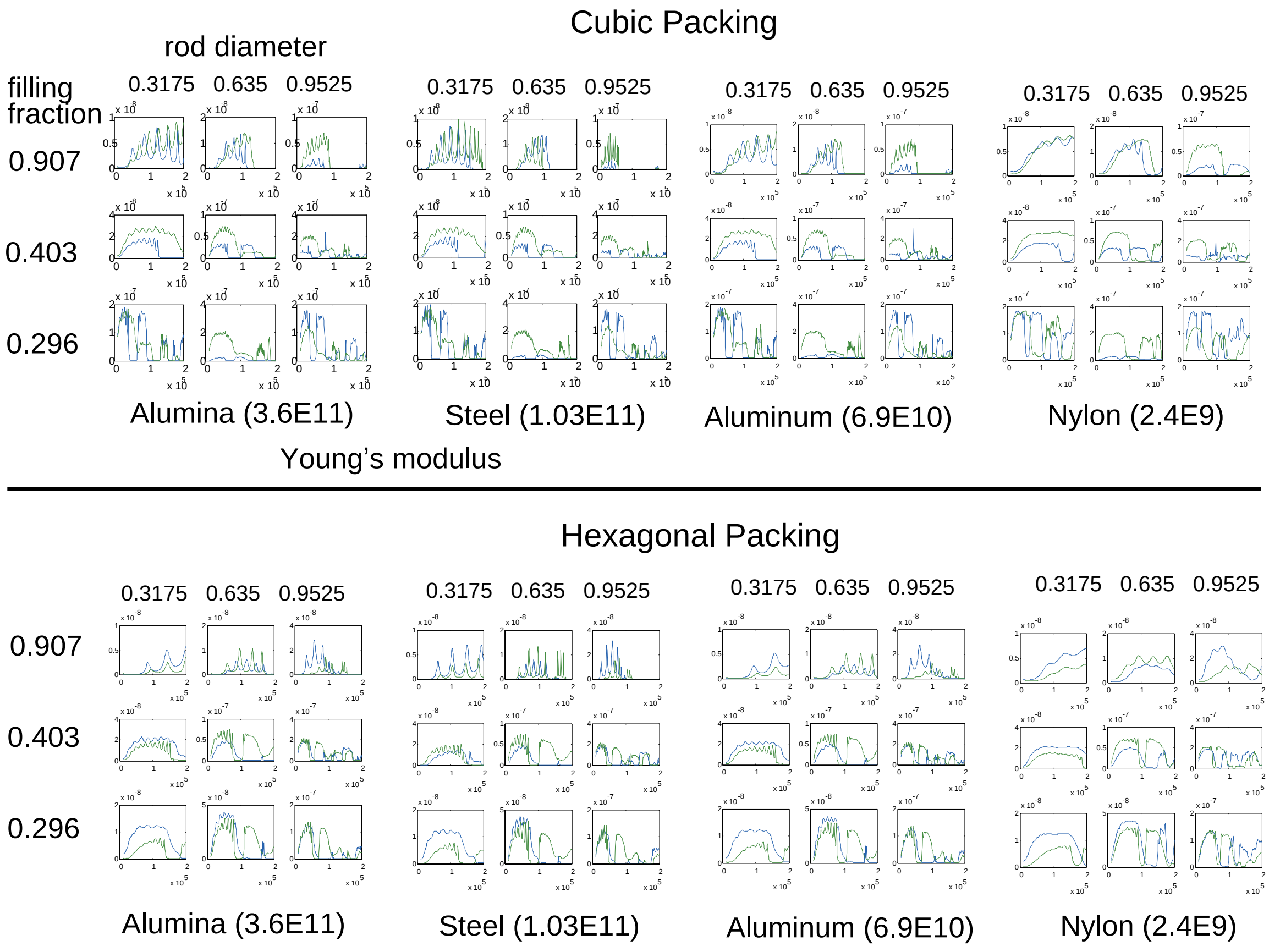

**Figure 2. Acoustic transmission spectra for a wide variety of materials and crystal types.**

These data can be summarized in two ways that both amplify the simple observations above. In Figure 3 we can plot the upper and lower edges of the first band-gap (the lowest frequency gap) as a function of modulus. Four sets of 3 X 3 blocks of graphs are produced from this. Each of the 3 X 3 consists of the filling fraction decreasing from left to right and the rod diameter increasing from top to bottom. The abscissa on each of the small graphs is the Young's modulus (log scale) and the ordinate is the frequency in kHz. Some of the 3 X 3 blocks have missing graphs because there was no definitive band gap (e.g. nylon filling fraction 0.4). Further, where the upper band-gap edge is at 200 kHz this simply means it was the limit from the simulations. We did not run the calculations above 200 kHz.

The first thing to notice about this series of graphs is the general trend of widening of the band-gap as the modulus increases. This rough trend is observed in many of the small graphs in the 3 X 3 blocks.

A final way to present these data makes use of further simplifications. If we measure the band gap width and center frequency and we average these for each of the rod diameters and further plot only the data for the lowest filling fraction, we get the graphs shown in Figure 4. The curves in this figure also include calculations for polyester and graphite. The graph on the upper left shows the center frequency in kHz for the first band-gap in cubic packed crystals with filling fraction of 0.296. The two curves are for the X and M directions and each data point represents the average of the three rod diameters. Except for one point (Young's modulus 4.4 GPa, polyester), we see two different trends, depending on the crystal orientation. In the X direction the center frequency tends to drop as the modulus increases, whereas in the M direction the center frequency tends to increase.

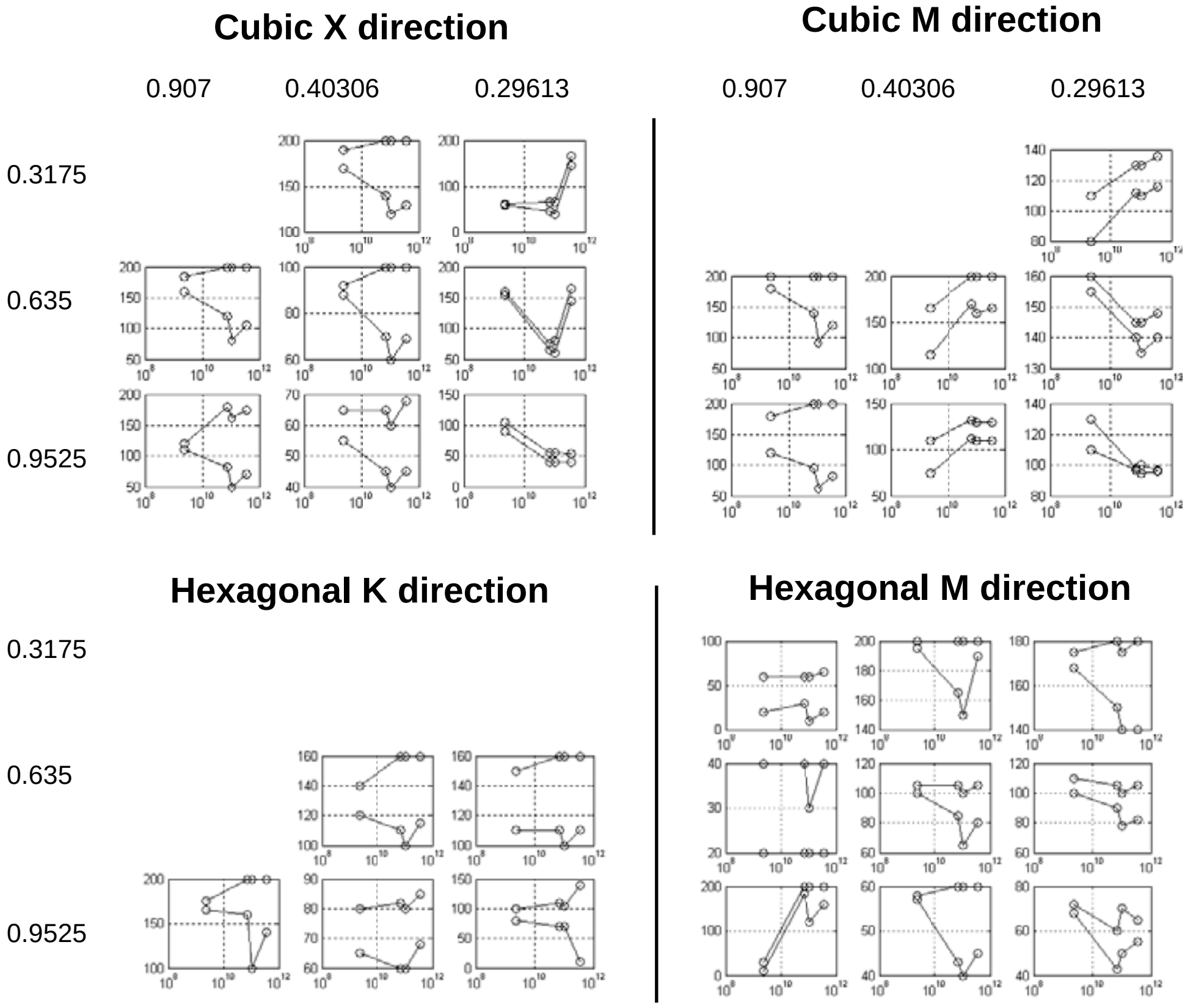

**Figure 3. Upper and lower ranges in frequency for the first band-gap in the observed spectra of Figure 2.**

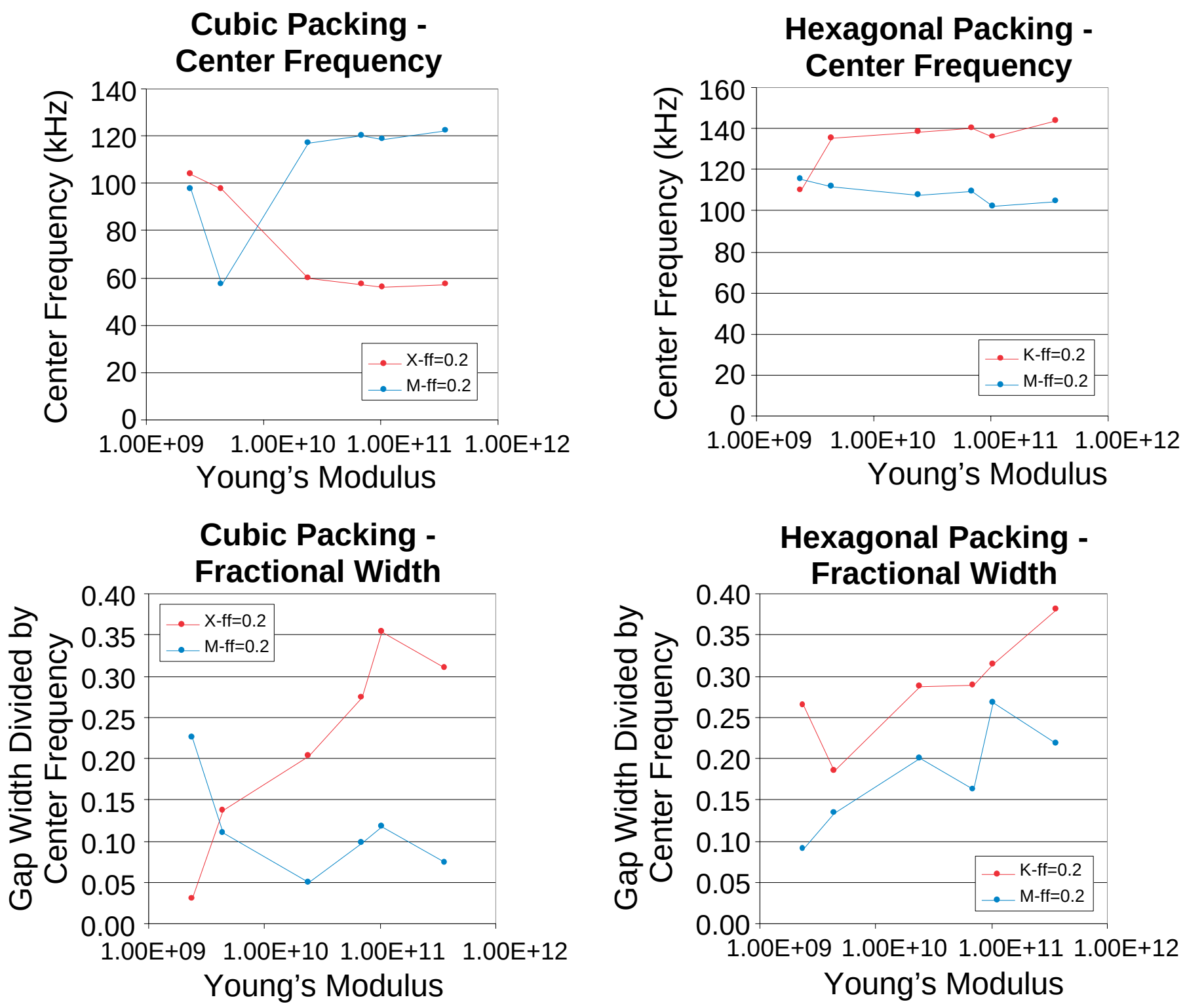

**Figure 4. Simplified graphical demonstration of the normalized gap width and the center frequency for 0.2 filling fraction and average of rod diameter.**

The lower left graph (cubic packing and fractional width of the first band-gap) shows the opposite trends between X and M direction. The upper right graph is for the hexagonal packing, and shows essential no variation in center frequency as a function of modulus, but indicates a shift of 30 kHz (increase) from the M direction to the K direction. The graph on the lower right shows essentially an increase in the gap width as Young's modulus increases in both K and M directions.

## Elementary Devices

In this section we quickly survey a few of the devices that are possible through band-gap engineering. The first example is for a machine mount to reduce acoustic vibration from engines and machines. The second example is frequency selective filtering in waveguides, and an acoustic lens.

From the above we see that phononic crystals are obviously useful for blocking sound at specific frequencies. Using them for acoustic signature reduction in machinery is an obvious application. Conventional machine mounts are essentially heavy springs. They typically have a flat performance between 500 Hz and 10 kHz and dampen signatures by about 20-30 dB. Below 500 Hz and above 10 kHz the performance is can be quite poor. The first design rule for phononic crystals is to select the size of the elements and the spacing between the elements to be comparable to the wavelength of sound you are interested in affecting. The center of the band-frequency will be on the order of $1/a$ where $a$ is the lattice spacing. Using Figure 3 as our design rules, the first thing we select is two materials with widely differing acoustic impedance: steel and foamed steel. The foamed metal will have an acoustic impedance similar to air.

We also see from our graphic results that the hexagonal lattice will in general give wider band gaps than square lattice. This is true only because our rods had a circular cross-section. By matching the symmetry of the elements comprising the crystal with the lattice symmetry in the Brillouin zone, the band-gap can be improved. Circular rods in a cubic lattice have poorer symmetry match than circular rods in a hexagonal lattice. So we can extrapolate and use cubic elements to build our cubic packed phononic crystal to achieve wider band-gap. These issues of element shape and lattice type were discussed in Sliwa and Krawczyk[18] who support our conjecture.

Our phononic crystal machine mount consists of cubic (4 cm edge) steel and 4 cm cubic foamed steel. These cubes were packed in cubic lattice with 2 mm gap. The structure consists of two layers of steel cubes followed by a layer of foamed cubes. This three-layer structure was repeated three times for the final machine mount. Figure 5 shows the structure and the acoustic transmission spectra for the device. By using three layers of the three layers we can enhance the performance. When the advancing sound front meets the solid steel, the phase will speed up. When the advancing sound front meets the foamed steel, the phase will slow down. These two conflicting processes will add to the constructive and destructive interference of the elastic waves. When the advancing wave front meets the foamed steel layer it will be reflected at some frequencies this reflected wave will phase match with the forward wave giving constructive interference. The constructive interference waves will not be transmitted. Only the destructive interference waves will be transmitted.

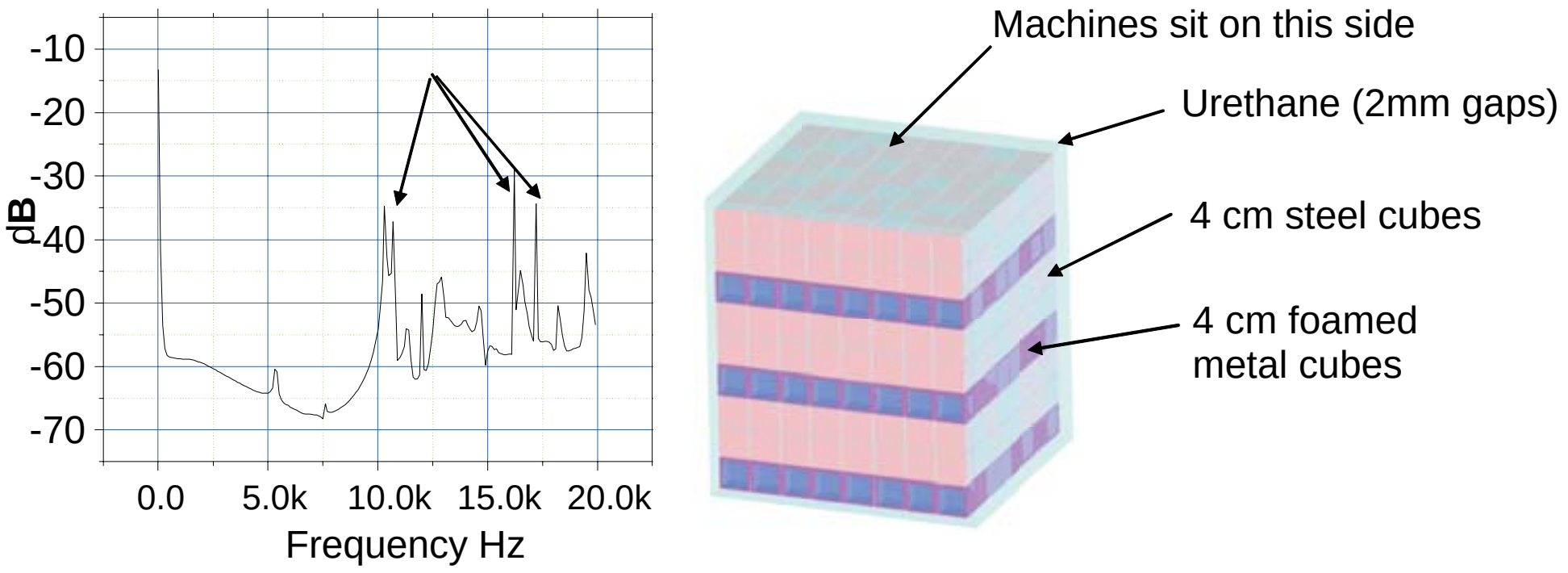

**Figure 5. Improved machine mount for acoustic signature reduction and transmission spectra from 0 Hz to 200 kHz.**

Though our machine mount is a little thicker than the big springs used for some heavy machines, this should not hinder its application in some domains. The machine mount shown in Figure 5 is 37.6 cm in height. As seen from the transmission spectra, the phononic machine mount rapidly drops the acoustic signature below 60 dB from about 20 Hz to 10 kHz. This is already an improvement of 20-30 dB over the state-of-the-art. The transmission spectra above 10 kHz show some sharp resonances (indicated by arrows in the figure) slightly above -40 dB. These can be reduced below -50 dB by slightly misaligning the layers of steel cubes.

As a second application we examine a frequency selective waveguide or wavelength division multiplexing (WDM). In phononic crystals the frequencies within the band-gap are reflected and do not get through the crystal. By creating a channel (or waveguide) of defects in the lattice we can direct specific frequencies in the band-gap along the waveguide, thus allowing them to pass unhindered only in the waveguide. We demonstrate this waveguide concept and further exploit the evanescence of waves in the waveguide to show how the acoustic waves can couple and propagate in a third waveguide.

The phononic crystal was created with aluminum rods of 3.0 cm diameter and packed in a hexagonal lattice with a filling fraction of 0.5804. The rods were surrounded by urethane impedance matched to water. The crystal is 0.7 X 0.7 m with the pressure source on the left side of the crystal and the other boundaries set for urethane impedance. Figure 6 shows the transmission spectra for the phononic crystal. Notice there is a wide band-gap centered at about 40 kHz.

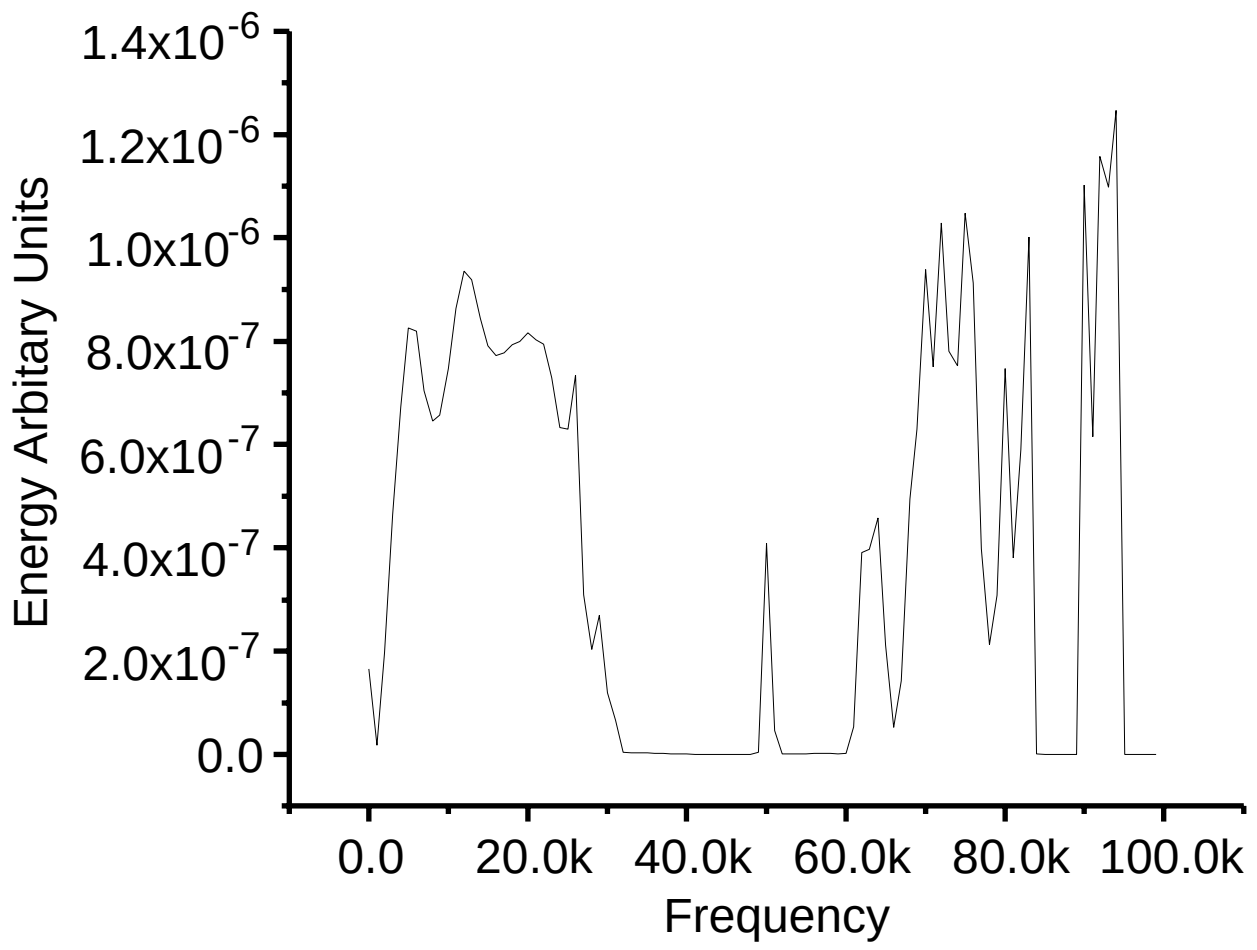

**Figure 6. Acoustic transmission spectra for phononic crystal used to study waveguides. This spectrum is prior to creating the channel defects in the lattice.**

Figure 7 shows the acoustic spectra inside the channels. The figure also shows pressure plots at one time instant. The spectra show resonances in the channels. At 45.450 kHz the acoustic signal travels through the two side channels and at 47.450 kHz the signal travels through the straight channel. Notice that the straight channel has four rods plugging the entrance at the end with the acoustic source and that the two side channels are open to the source. At 47.450 kHz the signal transfers from the two outer channels to the central inner channel. The evanescence of the waves in the two waveguides join through a beat phenomenon and transfer into the central waveguide. The end result is that the arrangement of waveguides can be used for multiplexing of acoustic signals.

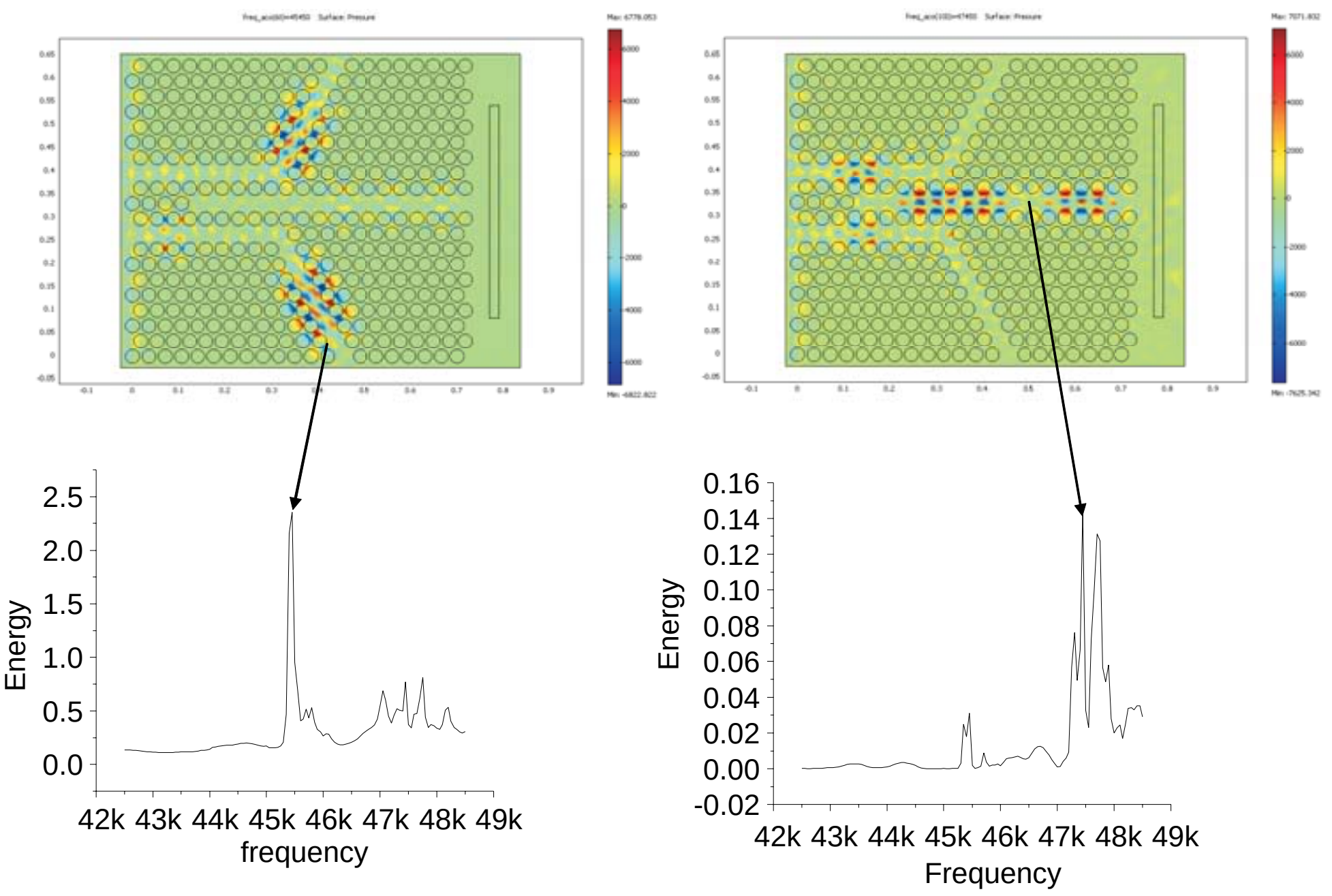

**Figure 7. Frequency selective waveguides and their respective spectra in side the channel.**

A further example using the same crystal (3.0 cm Al rods, 0.5804 filling fraction, hexagonal packing) but this time packed in air instead of urethane we created a lens by removing elements. Since the phase of the acoustic wave will speed up or slow down depending on the impedance of the elements and arrangement and number of elements in its path we can simply remove rods to form a converging lens. The lens we created resembles a converging lens from optics. Figure 8 shows the lens and its performance at 4 kHz. Clearly there is a focal point. Integration of the acoustic energy at the focal region and above it show differences in the transmission spectra from 3 kHz to 6 kHz.

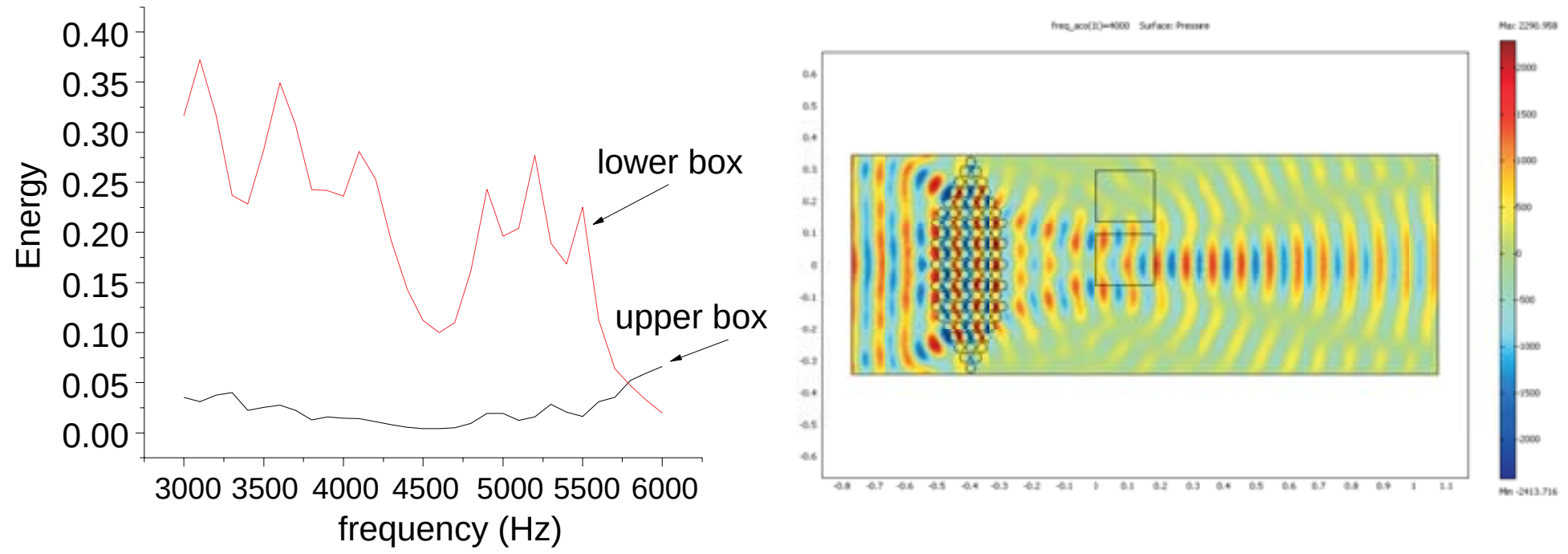

**Figure 8. Acoustic lens created from a phononic crystal. The two spectra are for the transmitted energy at the two locations indicated in the pressure plot of the lens.**

## Conclusions

We have done an extensive computational survey of crystal packing arrangements, filling fraction, material type and crystal orientation. Obviously the most important consideration for band-gap engineering is the size of the elements comprising the crystal and the spacing of the elements. In general the center of the band-gap is proportional to the inverse of the wavelength and the inverse of the lattice spacing: $\omega_g \sim 1/\lambda \sim 1/a$.

Though our study was limited to only three filling fractions and three rod diameters, Kushwaha and Helevi[3] have shown that gap width follows a parabolic relation with filling fraction. At low filling fraction the gap center increases and reaches a maximum where filling fraction is about 0.35. Above that, the gap center decreases. This makes intuitive sense because at some threshold in filling fraction, the impedance is dominated by one material and shifts the other material.

A second rule of thumb is that the impedance mismatch is very important. Contrary to some reported work, the mass density is not the next most important parameter. For example, lead has a density of 11,200 kg/m$^3$ and a Young's modulus of 5.42 X 10$^{10}$ Pa while the comparable values for aluminum are 3860 kg/m$^3$ and 6.9 X 10$^{10}$ Pa. The difference in modulus is greater than one GPa. This is in agreement with Newton's conjecture of elastic wave propagation through media. Our work clearly shows that the second most important consideration for acoustic band-gap engineering is selecting materials with appropriate Young's modulus. The gap width tends to increase as Young's modulus increases.

Another rule of thumb is that for maximum gap width the shape of the elements in the crystal should be closely matched to crystal packing. By matching the symmetry of the elements with the symmetry in the Brillouin zone the gap can be maximized. Clearly in our work hexagonal packing with round rods are better symmetry-matched than round rods in cubic packing. This has also been confirmed by Sliwa and Krawczyk.[18]

Using some of the above rules of thumb for band-gap engineering we presented some simple application ideas. Because of the ability of phononic crystals to filter sound at specific frequencies an obvious application is machine mounts. We designed a machine mount that performs 20 to 30 dB better than state-of-the-art in acoustic signature reduction.

Beyond acoustic filtering applications, the waveguide multiplexing and the lens suggests analogies with photonic crystals. It is interesting to speculate that if photonic crystals are being pursued for photonic computing than perhaps phononic crystals could be used for acoustic computing. Large blocks of engineered systems could potentially manipulate acoustic signals like integrated optical devices manipulate light.

Though we have described some rules of thumb for acoustic band-gap engineering in practice, because of the large number of adjustable parameters, it is probably more cost effective to use some computational guided-search to match a specific phononic crystal architecture with an application. Sigmund and Jensen[5] and Hakansson and Sanchez-Dehesa[13] use genetic algorithms for phononic crystal design. It would be interesting to use genetic algorithms for designing integrated acoustic signal processing systems similar to integrated optics.

**Acknowledgments**

We thank G. E. Caledonia and B. D. Green for helpful conversations; and we thank W. J. Marinelli for technical discussions and preparation of Figure 4.